\documentstyle[preprint,aps,prl]{revtex}
\input epsf
\begin{document}
\draft

\title{On Equivalence of Critical Collapse of Non-Abelian Fields}

\author{Piotr Bizo\'n${}^{1}$, Tadeusz Chmaj${}^{2}$, and 
Zbis\l aw Tabor${}^{1}$}
\address{$^{1}$Insitute of Physics, Jagiellonian University, Cracow,
 Poland}
\address{$^{2}$Institute of Nuclear Physics, Cracow, Poland}
\date{November 28, 1998}

\maketitle

\begin{abstract}
We continue our study of  the gravitational collapse of spherically symmetric
skyrmions. For certain families of initial data, we find the
discretely self-similar Type II critical transition characterized
by the mass scaling exponent $\gamma \approx 0.20$ and the echoing
period $\Delta \approx 0.74$. We argue that the coincidence of these
 critical exponents with those found previously in the
Einstein-Yang-Mills model is not accidental but, in fact, the two 
models belong to the same universality class.
\end{abstract}
\pacs{04.25.Dm, 04.40.-b, 04.70.Bw}


This is a second paper in the series devoted  to the study of critical
gravitational collapse in the Einstein-Skyrme (ES) model. The main
motivation for using this model is an attempt to understand the role
of different length scales and stationary solutions in the dynamics of
Einstein's equations at the treshold of black hole formation.
 In the first paper~\cite{my} we showed  that the presence of
sphaleron solutions gives rise to  Type I critical behavior for
certain families of initial data. In particular, we found a new kind
of first order phase transition in which  subcritical data relax to the
static gravitating skyrmion.
 Here, we focus our attention on Type II
critical behavior. After providing the numerical evidence for the
existence of such transition, we discuss the universality
of critical behavior with respect to intrinsic scales present in the model.
We show that the  two length scales become gradually
irrelevant on small spacetime scales near 
criticality and
consequently  the near-critical solutions can be asymptotically
self-similar. Interestingly enough, the critical exponents
characterizing
the Type II transition, the mass-scaling exponent $\gamma \approx 0.20$ and
the echoing period $\Delta \approx 0.74$, agree (within the numerical error
limits)
 with those previously
found in the Einstein-Yang-Mills (EYM) model~\cite{eym}. We
argue that this fact is a reflection of  equivalence of Type II
critical transitions between the  EYM and the ES models, that is the
critical solutions in both models are asymptotically identical.
To our knowledge, this is the first example of universality of
critical  collapse 
across two physically fundamentally   different systems (the
 universality classes
observed previously in collapse simulations comprised models differing
only by potential terms in  a lagrangian).

For this paper to be self-contained let us briefly recall the setup for the
spherically symmetric ES model we used in~\cite{my}. Adopting the 
polar time slicing and the areal radial coordinate, the spacetime
metric
can be written as
\begin{eqnarray}
ds^2 &=& -N(r,t) e^{-2\delta (r,t)} dt^2 + N^{-1} (r,t) dr^2 + r^2 d \Omega^2.
\label{METRIC}
\end{eqnarray}

As a matter source we take the $SU(2)$-valued scalar field $U(x)$ (called the
chiral field) and 
 assume the hedgehog ansatz $U=exp(i\vec\sigma \cdot
\hat r F(r,t))$, where $\vec\sigma$ is the vector of Pauli matrices.
The components of stress-energy tensor $T_{ab}$ expressed in the orthonormal
 frame determined by the metric (1) are
\begin{equation}
T_{00} = \frac{u}{2 r^2} (N F'^2+N^{-1} e^{2\delta} \dot F^2)
+ \frac{\sin^2{F}}{r^2} \left(f^2+\frac{\sin^2{F}}{2 e^2 r^2}\right),
\end{equation}
\begin{equation}
T_{11} = \frac{u}{2 r^2} (N F'^2+N^{-1} e^{2\delta} \dot F^2)
- \frac{\sin^2{F}}{r^2} \left(f^2+\frac{\sin^2{F}}{2 e^2 r^2}\right),
\end{equation}
\begin{equation}
T_{01} = \frac{u}{r^2} e^{\delta} \dot F F',
\end{equation}
where overdots and primes 
 denote $\partial / \partial t$ and $\partial / \partial r$
 respectively, and $u=f^2 r^2+\frac{2}{e^2}\sin^2{F}$.
The two coupling constants $f^2$ and $e^2$ have dimensions:
$[f^2]=M L^{-1}$ and $[e^2]=M^{-1}L^{-1}$ (we use
units in which $c=1$). In order to write the evolution
 equations
in the first order form, we define an auxilary variable
$P=u e^{\delta} N^{-1} {\dot F}$.
Then, the full set of  ES equations is
\begin{eqnarray}
\!{\dot F} &=& e^{-\delta} N \frac{P}{u},
\label{fDOT}
\\
\!{\dot P} &=& (e^{-\delta} N u F')' \!
- \!\sin(2F) e^{-\delta} \left[f^2+\frac{1}{e^2}\left(
  \!N F'^2 -\! N \frac{P^2}{u^2}
+ \!\frac{\sin^2{\!F}}{r^2}\right)\right],
\\
N' &=& \frac{1-N}{r} - 8 \pi G r\,T_{00},
\\
{\dot N} &=& -8 \pi G r e^{-\delta} N \,T_{01},
\\
{\delta'} &=& -4 \pi G r N^{-1} (T_{00}+T_{11}).
\end{eqnarray}
In order to make an identification of certain terms in the equations easier,
all the coupling constants are displayed, however we remind  that,
 apart from the overall scale, solutions depend on
these constants only through the dimensionless parameter $\alpha=4\pi
G f^2$. We solve
Eqs. (5-9)  for regular asymptotically flat initial
data.
 The condition of  regularity at the center $N(r,t)=1+O(r^2)$ is
ensured by the boundary condition 
$F(r,t)=O(r)$ for $r \rightarrow 0$. The asymptotic flatness 
of initial data $N(r,0)=1+O(1/r)$ for $r \rightarrow \infty$ is
guaranteed 
by the initial condition
$F(r,0)=B \pi + O(1/r^2)$,
where the integer $B$, usually referred to as the baryon number, is
the  topological
degree of the chiral field. Since the baryon number is dynamically preserved,
the Cauchy problem falls into infinitely many superselection sectors
labelled by $B$ - here we concentrate on the $B=0$ sector.
 A typical one-parameter family of
initial data
in this class, interpolating between black-hole and no-black-hole 
spacetimes, is
an initially ingoing ``Gaussian''
\begin{equation}
F(r,0)= A\, r^3 \exp\left[-\left(\frac{r-r_0}{s}\right)^4\right],
\end{equation}
where one of the parameters $A, s$, or $r_0$ is varied (hereafter this
parameter is denoted by $p$) while the
others 
are fixed. As usually, the
critical value $p^{*}$ is located by performing a bisecting search in
$p$. In order to get into close proximity of $p^{*}$ we have implemented an
 adaptive mesh refinement 
algorithm. This code was essential in probing the critical region with
 sufficient  resolution.
Our numerical results demonstrate the existence of Type II
critical transition with its two main characteristic features, first
observed by Choptuik in the massless scalar field
collapse~\cite{matt}, namely:
\begin{itemize}
\item{Mass scaling: For supercritical data, the final black hole mass
scales as 
$m_{BH} \sim
 C |p-p^{*}|^{\gamma}$ with the exponent $\gamma \approx 0.20$. As
 shown in Fig.~1 this scaling law holds over two orders of magnitude
 of mass.}
\item{Echoing:  For near-critical data, the solutions approach
(for sufficiently small $r$) a certain universal intermediate
attractor which is discretely self-similar with the echoing period
  $\Delta \approx 0.74$. This is illustrated in Fig.~2.}
\end{itemize}

We find that the critical exponents, $\gamma$ and $\Delta$, are
 universal not only
with respect to initial data, but also with respect to the parameter
 $\alpha$. 
The universality of Type II critical collapse with respect to a
dimensionless parameter is a rare phenomenon which can occur only
if the parameter enters evolution equations through terms which become
 ``irrelevant'' near criticality
(see the discussion of this issue in Gundlach's
 review~\cite{gundlach1}).
 Now, we
 would like to show that this is exactly what happens for
$\alpha$ in the ES equations. Along the way, we will
 see that discrete self-similarity is compatible with the
 presence of two length scales in the model. Our basic assumption is
 that $F(r,t)/\sqrt{r}$ is an echoing quantity (scaling variable).
 This assumption is not
 only justified empirically (see Fig.~2) but, as
follows from a simple dimensional analysis of Einstein's equations
 (7-9),
 it seems to be the only possibility compatible with
the discrete self-similarity of metric functions $N$ and $\delta$.
We  introduce a unit of length $L=\sqrt{4\pi G}/e$ and
dimensionless coordinates
\begin{equation}
\tau=\ln\left(\frac{t^{*}-t}{L}\right)
 \quad \mbox{and}\quad \xi=\ln\left(\frac{r}{t^{*}-t}\right)
\,,
\end{equation}
where $t^{*}$ is the accumulation time of the infinite number of echos.
We also define  new dimensionless variables 
\begin{equation}
\Phi(\tau,\xi)=\frac{F(r,t)}{\sqrt{r/L}}, \quad 
Z(\tau,\xi)=\sqrt{rL}\,F'(r,t), \quad
\Pi(\tau,\xi)=\sqrt{rL}\,e^{\delta}N^{-1} \dot F(r,t).
\end{equation}
By assumption $\Phi,Z$, and $\Pi$ are the scaling variables, that is
they are asymptotically periodic in $\tau$: $\Phi(\tau,\xi)\approx
\Phi(\tau+\Delta,\xi)$ etc. for large negative $\tau$ and fixed $\xi$.
Rewritting  Eqs.(5-9) in these new variables, we find that 
 $\alpha$ is always  multiplied by $e^{\tau}$, and the
only other terms depending explicitly on $\tau$ appear through the combination
\begin{equation}
X=e^{-\frac{1}{2}(\tau+\xi)} \sin\left(e^{\frac{1}{2}(\tau+\xi)}
  \Phi\right).
\end{equation}
For example, the hamiltonian constraint (7) takes the form
\begin{equation}
1-N-\frac{\partial N}{\partial \xi} =
N (\alpha e^{\tau+\xi}+2 X^2) (Z^2+\Pi^2) + X^2 (2 \alpha
e^{\tau+\xi}+ X^2).
\end{equation}
In the limit $\tau \rightarrow -\infty$, the terms
containing
$\alpha$ become negligible
(``irrelevant'' in the language of renormalization group theory), and 
therefore  the  critical behavior does not depend on $\alpha$.
In the same limit, 
$X \rightarrow \Phi$, so the equations become asymptotically
autonomous in $\tau$, and thereby scale invariant.
The equivalence of Type II critical behavior between ES and EYM models
is an immediate consequence of the universality with respect to
$\alpha$, because, as we pointed out in~\cite{my1}, for $\alpha=0$ 
Eqs.(5-9) reduce (after the substitution $w=\cos{F}$) to the EYM
equations.
This means that the EYM  critical solution constructed
by Gundlach~\cite{gundlach2} is valid (to the leading order) in the ES
case as well, and hence
 the critical exponents in both models are the same.  As mentioned
 above, this theoretical retrodiction is confirmed numerically.

We conclude with two remarks concerning possible extensions of the
research presented here and in~\cite{my}.
First, we would like to point out that by setting all terms in
Eqs.(5-9) containing the coupling constant $e^2$ (the Skyrme terms) to
zero,  one obtains
the $\sigma$-model coupled to gravity. This model is scale invariant,
so it does not admit a Type I transition but a Type II transition is
expected to exist. If so, the natural scaling variables will be
\begin{equation}
\tilde \Phi(\tau,\xi)=F(r,t), \quad 
\tilde Z(\tau,\xi)=r F'(r,t), \quad 
\tilde \Pi(\tau,\xi)= r e^{\delta}N^{-1} \dot F(r,t).
\end{equation}
 It is easy to check
that in this case the terms proportional to $\alpha$ are {\em not}
irrelevant. For example the analogue of Eq. (14) is
\begin{equation}
1-N-\frac{\partial N}{\partial \xi} =
\alpha N (\tilde Z^2+\tilde \Pi^2) + 2 \alpha \sin^2{\tilde \Phi}.
\end{equation}
Therefore, the critical solution, and {\em eo ipso} the critical exponents,
are anticipated to depend strongly on $\alpha$. We have been informed
that the group
of  researchers led by Peter Aichelburg is in the process of
investigating this and related problems~\cite{vienna}. If their
results confirm our expectation, the universality with respect to
$\alpha$ in the ES model could be interpreted as another
nonperturbative effect of the Skyrme correction to the $\sigma$-model.

Second, we recall that, in contrast to Type II, the Type I
critical transition in the ES model is manifestly nonuniversal with respect to
$\alpha$ because the critical solution (the sphaleron) changes with 
$\alpha$, in particular, it exists only for sufficiently
small $\alpha$. Thus, for large $\alpha$ only Type
II behavior is possible, while for small  $\alpha$ the two types of critical
behavior can coexist. In the latter case, one can anticipate the existence
of crossover effects at the
border of basins of attractions of Type I and Type II
critical solutions. We leave an investigation
 of these fascinating effects to other
researchers who, as we have recently accidentaly found
out~\cite{texas}, are also interested in the ES model. We would like
to emphasize that
a full  description of an extremely rich phenomenology of the ES model
is not a {\em per se} goal our studies -- for us this model is only
a testing ground for addressing  certain issues of the dynamics of
Einstein's equations in the presence of intrinsic scales and
 stationary (stable and unstable) solutions.

{\em Acknowledgments.} 
This research was supported  by the KBN grant
PB 99/P03/96/11. Part of the computations were performed on the
facilities of the
Institute
of Nuclear Physics in Cracow supported by the Stiftung f\"{u}r
Deutsche-Polnische Zusammenarbeit, project 1522/94/LN.

\begin{figure}
\epsfxsize=8in
\centerline{\epsffile{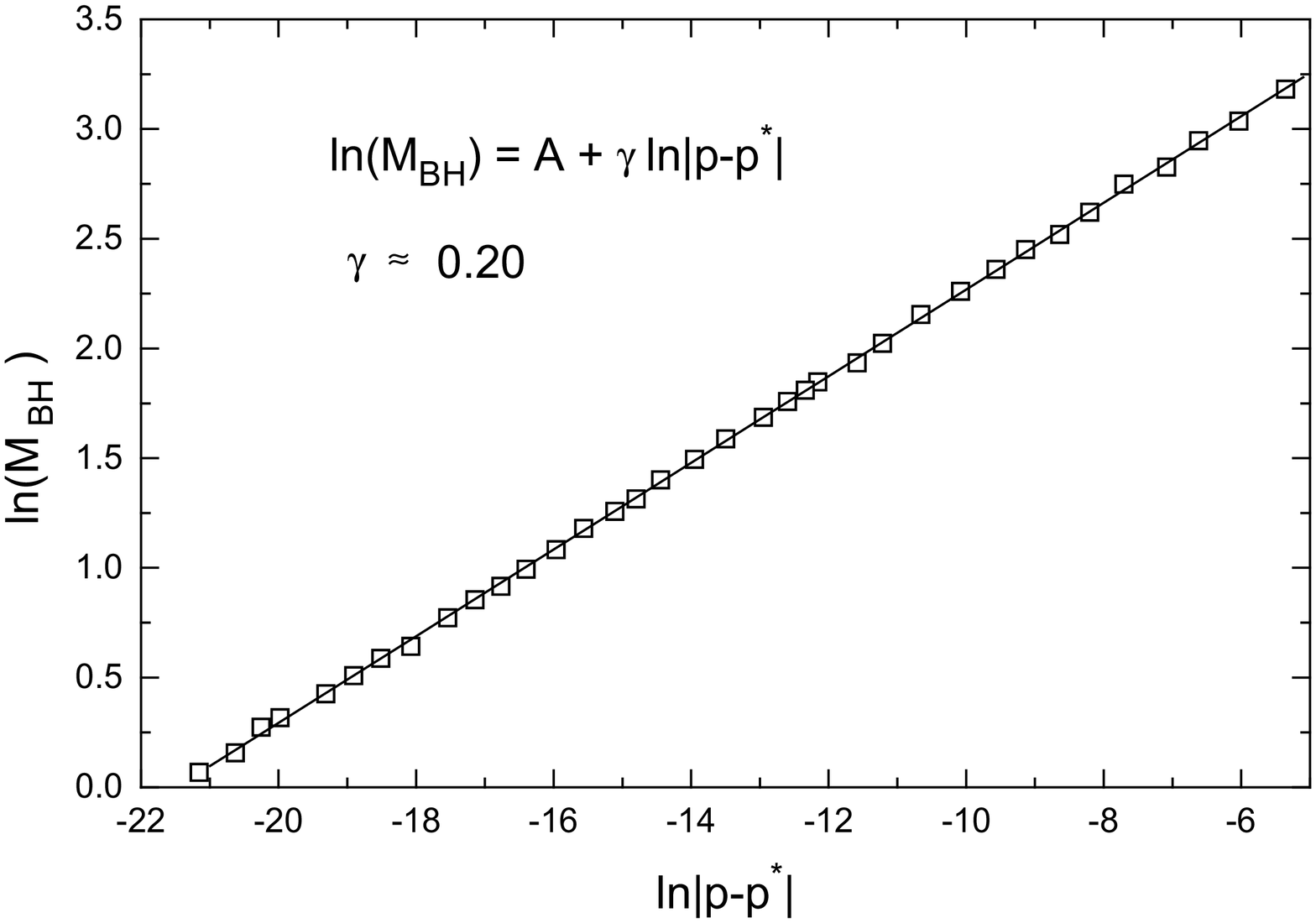}}
\caption{The black hole mass scaling. The logarithm of  black hole
mass $M_{BH}$ is plotted versus the logarithmic distance from
criticality
$\ln|p-p^{*}|$ for supercritical solutions generated from initial
data (10) for 
$\alpha=0.02$. The power-law fit
is indicated by the solid line with the slope $\gamma \approx 0.20$.}

\label{FIG1}
\end{figure}
\begin{figure}
\epsfxsize=8in
\centerline{\epsffile{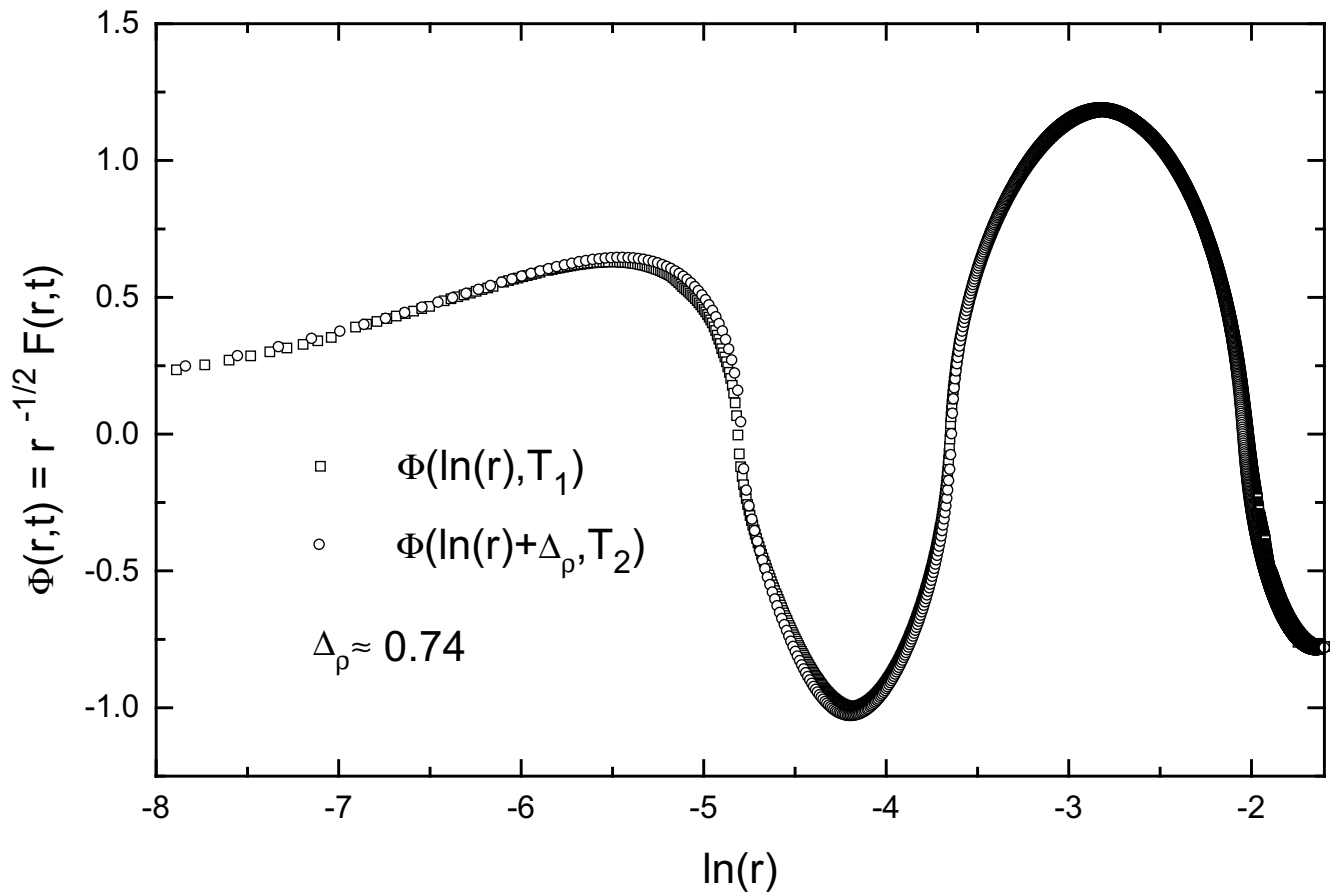}}
\caption{Numerical evidence for echoing. For a marginally critical
  solution generated from initial data (10) for $\alpha=0.02$ we plot 
the profile of $\Phi=F/\sqrt{r}$ versus
  $\rho=\ln(r)$ for some arbitrary late time $t_1$ and
  superimpose the profile of the first echo at time $t_2$ shifted by
$\rho \rightarrow \rho+\Delta_{\rho}$. The time $t_2$ and the radial
echoing period $\Delta_{\rho}$ are chosen to minimize the mean square
difference between the
two profiles. By repeating this calculation for a sequence of pairs
$(t_1,t_2)$, we estimated the temporal echoing  period $\Delta_{\tau}$
  from the slope of the line 
$t_2=t^{*}(1-e^{-\Delta_{\tau}})+e^{-\Delta_{\tau}} t_1$, confirming
  the
expectation that $\Delta_{\rho}=\Delta_{\tau}$.
}
\label{FIG2}
\end{figure}

\end{document}